\documentclass[twocolumn,english,aps,prd,longbibliography,groupedaddress]{revtex4-1}
\usepackage[T1]{fontenc}
\usepackage[utf8]{inputenc} 
\usepackage{color}
\usepackage{bm}
\usepackage[linesnumbered,ruled,vlined]{algorithm2e}
\usepackage{xcolor}
\usepackage{amsmath}
\usepackage{mathtools}
\usepackage{amssymb}
\usepackage{graphicx}
\usepackage{natbib}
\usepackage{braket}
\usepackage{psfrag}
\usepackage{tikz}
\usepackage[normalem]{ulem}
\usepackage{qcircuit}
\usepackage{siunitx}
\usepackage{placeins}
\usepackage[colorlinks=true,urlcolor=blue,citecolor=blue,linkcolor=blue]{hyperref}
\pdfmapfile{+sansmathaccent.map}

\usepackage{url}

\begin{document}
 

\title{Canonical Construction of Quantum Oracles}

 
\author{Austin Gilliam}
\affiliation{
JPMorgan Chase}

\author{Marco Pistoia}
\affiliation{
JPMorgan Chase}

\author{Constantin Gonciulea}
\affiliation{
JPMorgan Chase}
 
 
\date{\today}
 
\begin{abstract}
Selecting a set of basis states is a common task in quantum computing, in order to increase and/or evaluate their probabilities.
This is similar to designing \texttt{WHERE} clauses in classical database queries.
Even though one can find heuristic methods to achieve this, it is desirable to automate the process.
A common, but inefficient automation approach is to use oracles with classical evaluation of all the states at circuit design time.
In this paper, we present a novel, canonical way to produce a quantum oracle from an algebraic expression (in particular, an Ising model), that maps a set of selected states to the same value, coupled with a simple oracle that matches that particular value.
We also introduce a general form of the Grover iterate that standardizes this type of oracle.
We then apply this new methodology to particular cases of Ising Hamiltonians that model the zero-sum subset problem and the computation of Fibonacci numbers.
In addition, this paper presents experimental results obtained on real quantum hardware, the new Honeywell computer based on trapped-ion technology with quantum volume 64.
\end{abstract}
 
\maketitle
 

\section{\label{sec:intro}Introduction}
Let us take the simple example of a two-qubit quantum state described in one of the following two ways:

\begin{enumerate}
    \item $\ket{00}$ and $\ket{11}$ are equally possible, while $\ket{01}$ and $\ket{10}$ are impossible
    \item The measurements of the two qubits match
\end{enumerate}
Both descriptions impose constraints on the measurement outcomes of a quantum state, and match one of the Bell States~\cite{Bell1964},
$\frac{1}{\sqrt{2}}(\ket{00} + \ket{11})$.

In this paper, we analyze current strategies to impose and verify such constraints. Furthermore, we introduce a novel canonical oracle encoding methodology to efficiently represent algebraic constraints as partition functions. This approach can be seen as encoding a random variable instead of just a probability distribution, and allows for standardizing subsequent quantum computations involving operations such as searching and counting. 
The method is based on the Quantum Dictionary~\cite{Gonciulea2019, Gilliam2019}, whose original intent was to efficiently encode a given function into a quantum state for optimization purposes.
We contributed a version that uses adaptive search for Quadratic Unconstrained Binary Optimization (QUBO) problems to Qiskit~\cite{qiskit}. 
Elaborating on this methodology, this paper makes the following novel contributions:

\begin{enumerate}
    \item A practical demonstration that the proposed canonical oracle encoding methodology can be applied to NP-hard combinatorial computations.  We choose the zero-sum subset problem as an example.
    \item A generalized form of the Grover iterate~\cite{Grover1996, Brassard2000}, that standardizes oracle construction from simple matching oracles.
    \item A comparison between heuristic encoding, naive oracle encoding, and the proposed canonical oracle encoding methodology.
    We demonstrate these three types of encoding in the context of calculating Fibonacci numbers.
    \item A connection between the Amplitude Estimation algorithm and the generalized Born Rule, with the probability function of the probability space defined by a quantum computation and the probability mass function of a random variable.
    \item An experimental validation of the proposed methodology on real quantum hardware, the Honeywell System Model HØ quantum computer with quantum volume 64~\cite{QuantumVolume}.
\end{enumerate}
The remainder of this paper is organized as follows: Section~\ref{sec:properties} provides an overview of existing techniques for selecting and evaluating quantum states.
Section~\ref{sec:smart-oracles} introduces a new generalized form of oracles that can be used in Amplitude Amplification and Estimation.
Section \ref{sec:strategies} presents our novel canonical oracle encoding and compares it to two existing property-encoding strategies.
Section \ref{sec:examples} validates, both in theory and practice, (a) how the proposed canonical oracle encoding methodology can be applied to NP-Hard problems, with the zero-sum subset problem used as an example, and (b) an application of the three encoding strategies to another combinatorial problem---the computation of Fibonacci numbers. 
Experimental results are provided on real quantum hardware.
Finally, Section \ref{sec:conclusion} concludes the paper and discusses future work.

\section{\label{sec:properties}Selecting and Evaluating Quantum States}

Given a quantum computation, we are often interested in \emph{evaluating} a set of \emph{selected outcomes}, i.e., basis states satisfying a given property. In the remainder of this paper, we will interchangeably use the terms \emph{selected outcomes} and \emph{marked states} depending on the context. Marked states are sometimes called \emph{good states}, and the rest \emph{bad states}~\cite{Brassard2000}.

In particular, we are interested in answering the following questions:

\begin{enumerate}
    \item What is the probability of a single state?
    \item What is the probability of a set of states satisfying a property?
    \item How many states satisfy a certain property?
    \item How many states have a non-zero amplitude? 
    \item How many states have 0 (or another value) in a register? 
\end{enumerate}

The \emph{selection} of the desired outcomes (i.e., marking the corresponding basis states in the computational basis) typically requires an oracle that flips the phase of the basis states satisfying a certain property.
Some algorithms require the mapping of the good states to the $\ket{1}$ state of an \emph{oracle qubit}, without specifying how the mapping is done.
We show here that the Quantum Dictionary pattern~\cite{Gonciulea2019} can be used to provide a quantum implementation of such a mapping. 

The evaluation of the selected outcomes can be done classically or as part of the quantum computation.

When the evaluation of the selected outcomes is done classically, it is necessary to perform the computations a number of times, thereby resorting to classical sampling. The quantum computation serves only as a probability distribution, which can be as simple as the uniform distribution created by an equal superposition.

More interestingly, the evaluation of the  selected outcomes can be made part of the quantum computation.  For example, the probability of the set of outcomes can be computed using the Amplitude Estimation algorithm, and their number can be estimated using the Quantum Counting algorithm~\cite{Brassard2000} (see Appendix~\ref{sec:amplitude-estimation} for more details). These methods are based on a repeated application of the Grover iterate~\cite{Grover1996}, which in turn relies on an \emph{oracle} that wraps the property describing the desired outcomes.

\section{\label{sec:smart-oracles}Amplitude Amplification and Estimation with Generalized Oracles}
The Quantum Dictionary pattern allows for encoding key/value pairs in two entangled registers~\cite{Gonciulea2019}. In particular, keys can be used as indices of array values, or inputs of a (total) function. Polynomial functions can be encoded in a particularly efficient way, as described in~\cite{Gilliam2019}.

The crucial insight for using the Quantum Dictionary pattern to select quantum states is that if these states satisfy a certain algebraic equation, they can all be mapped to a single value of a function. Then, we can use a simple binary matching oracle on the value register to mark the selected inputs in the first register. The function encoding becomes part of the marking process, and therefore can be thought of as an enhancement of the oracle that simply matches a value. A self-contained introduction of the Quantum Dictionary can be found in Appendix~\ref{sec:quantum_dictionary}.

We consider this to be a generalization of the Amplitude Amplification algorithm~\cite{Grover1996, Brassard2000}, with the following ingredients:

\begin{itemize}
    \item Any unitary operator $A$, which creates a superposition state
    \item A \emph{property-encoding operator} $B$ that maps the selected states to a single function value
    \item A \emph{simple, binary-matching oracle} $O_B$, whose purpose is to match the value assigned to the selected states
    \item The \emph{diffusion operator} $D$, which multiplies the amplitude of the $\ket{0}_n$ state by $-1$
\end{itemize}
Based on the above, the Grover iterate is now defined as $G = ADA^{\dagger}O$ (in some contexts it is useful to add a negative in front), where $O=B^{\dagger}O_BB$ is the \emph{canonical oracle} that, by construction, flips the phases of the selected states, and the Amplitude Amplification routine takes the form of $BG^rA$, where $r$ is the number of times the Grover iterate is applied.

In the original version of Amplitude Amplification, the $B$ operator is not present.  When adding it, $B$ could be made integral part of $A$. However, since it is specific to selecting states, it makes more sense to integrate $B$ into a more complex oracle construct.
This enhanced form of the Grover iterate can be used in the Amplitude Estimation algorithm as well, leading to a generalized version of the algorithm.

To encode a function, the Quantum Dictionary pattern uses an $A$ operator that creates an equal superposition, with $B$ implementing a partition function that maps the desired inputs to a single value. We will provide examples of such partition functions in the following sections.

\section{\label{sec:strategies}Property-Encoding Strategies}
In this section, we define the three main strategies for encoding properties of selected outcomes for quantum computations.
The selection is made for either making selected outcomes stand out in measurements, or evaluating them in some way (typically probability estimation or counting).
In order to build the intuition for these strategies we will use the example of a simple Bell state.

\subsection{\label{subsec:amplitude-encoding}Heuristic Encoding}

\begin{figure}[ht]
\begin{equation*}
    \Qcircuit @C=1.0em @R=0.0em @!R {
	 	\lstick{\ket{0}} & \gate{H} & \ctrl{1} & \qw \\
	 	\lstick{\ket{0}} & \qw & \targ & \qw \\
	 }
\end{equation*}
\caption{\label{fig:bell-state-circuit}Bell State Circuit} 
\end{figure}
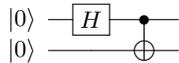

One way to have selected outcomes stand out in measurements is to make all basis states that do not satisfy the property impossible, i.e. by rendering their amplitude zero. In general, this is done heuristically, and is hard to automate. As an example, let us consider the Bell state, discussed previously, which can be prepared using the circuit in Figure~\ref{fig:bell-state-circuit}.

Once prepared, the quantum state can be classically evaluated by running the circuit a number of times and building a histogram from the outcomes. A run is usually called a \emph{shot}.

\begin{figure}[ht]
\includegraphics[width=8cm]{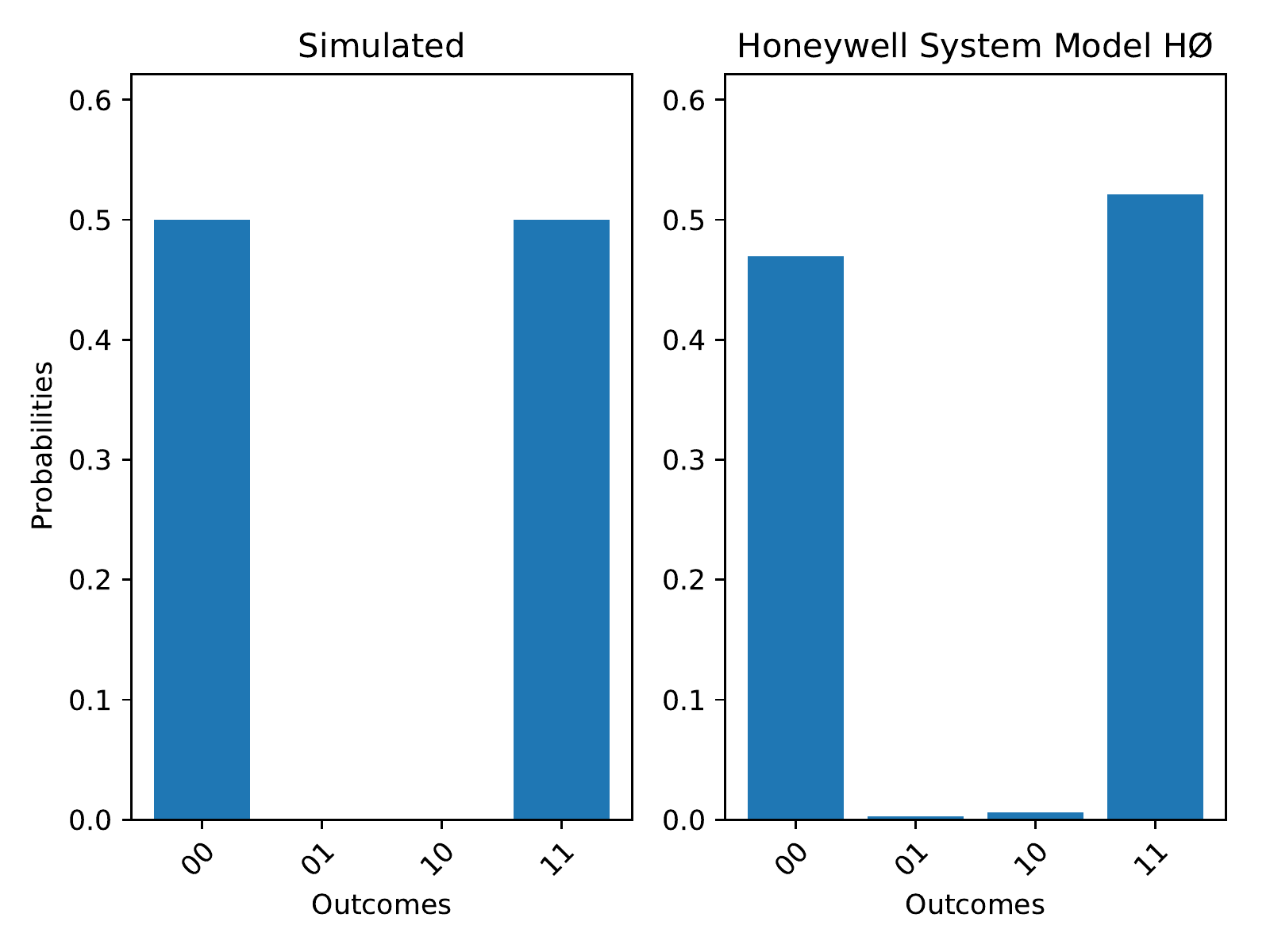}
\caption{\label{fig:bell-state-example}The result of running the circuit from Figure~\ref{fig:bell-state-circuit} on a simulator (left) and the Honeywell System Model HØ (right).}
\end{figure}

The corresponding probability distribution is shown in Figure~\ref{fig:bell-state-example}.
Note that when running on a real computer, even when the error rates are minimal, the theoretically-impossible outcomes will be measured a small number of times.
This is similar to classical computing, where numerical calculations are always an approximation.

\subsection{\label{subsec:oracle-encoding}Naive Oracle Encoding}
Unlike the heuristic approach, oracles allow the automation of the process of making the selected states stand out in measurements.   An oracle uses a predicate that is based on the property we want to encode, and is applied to all the basis states.  When the oracle's circuit is designed, for each basis state that satisfies the predicate, we add a quantum subcircuit that matches that particular state and multiplies its amplitude by $-1$.

Oracles are used in the Amplitude Estimation algorithm to approximately calculate the probability of a set of selected outcomes. For example, suppose we want to find the amplitude of a single outcome of the circuit that encodes the Bell State described above, e.g. $\ket{00}$. If we run the Amplitude Estimation algorithm with an oracle whose predicate matches the single state $\ket{00}$, using a simulator and 2 qubits for the result, the probability is $0.49999$, aligned with the expected value. Doing the same for $\ket{01}$ leads to probability $0.0$. 

\begin{figure}[ht]
\begin{center}
\begin{tabular}{ |c|c|c|c| } 
\hline
 marked state & expected & actual \\ 
 \hline
 $\ket{00}$ & 0.5 & 0.49999 \\ 
 $\ket{01}$ & 0.0 & 0.0 \\ 
 \hline
\end{tabular}
\end{center}
\caption{\label{fig:single-amplitude-ae-table}Expected and simulated results of an Amplitude Estimation circuit applied to a Bell State with the listed single-outcome property.} 
\end{figure} 

Now, suppose we want to find the probability of a set of outcomes satisfying a certain property. For example, let us consider the outcomes having the same digit in both positions, consisting of $\ket{00}$ and $\ket{11}$. Using an oracle with an appropriate predicate, if we run the Amplitude Estimation algorithm, the result is $1.0$. Doing the same for the outcomes having different digits, consisting of $\ket{01}$ and $\ket{10}$, the result is $0.0$. 

\begin{figure}[ht]
\begin{center}
\begin{tabular}{ |c|c|c|c| } 
\hline
 marked states & expected & actual \\ 
 \hline
 $\ket{00}$, $\ket{11}$ & 1.0 & 0.99998 \\ 
 $\ket{01}$, $\ket{10}$ & 0.0 & 0.0 \\ 
 \hline
\end{tabular}
\end{center}
\caption{\label{fig:muulti-amplitude-ae-table}The expected and simulated results of an Amplitude Estimation circuit applied to a Bell State with the listed multi-outcome property and precision.} 
\end{figure} 

As we can see, the Amplitude Estimation procedure obeys the sum rule: the estimated probability of a set of outcomes is the sum of the estimated probabilities of the single outcomes. This is a profound insight, as it aligns with the Generalized Born Rule and the measurement postulate of Quantum Theory that make the measurements of a quantum computation independent of each other. For more details, see Appendix~\ref{sec:amplitude-estimation}.

\subsection{\label{subsec:state-encoding}Canonical Oracle Encoding}
In our early research efforts the naive oracle encoding was useful, as it allowed for a level of automation, but it is generally inefficient due to the need to check the predicate on all basis states.

The canonical oracle encoding, introduced in this section, allows for efficient encoding of properties that can be expressed algebraically as polynomials of binary variables.  In particular, quadratic Hamiltonians, such as Ising models, lend themselves to this approach.

This methodology relies on values being associated to all outcomes in such a way that those outcomes satisfying a given property map to the same value, which can be subsequently used in algorithms such as Amplitude Amplification, Amplitude Estimation, and Quantum Counting.

Given an $n$-qubit quantum system and a list of integers $(a_0, a_1, ..., a_{n-1})$, we can mathematically represent the sum of the digits of a possible outcome (represented as a binary string of length $n$) as follows:

\begin{eqnarray}
S(x_0, x_1, ..., x_{n-1}) = \sum_{i=0}^{n-1}  a_ix_i
\label{eq:sum-of-digits}
\end{eqnarray}
where $x_i \in \{0,1\}$ for $i \in \{0, \ldots, n-1\}$.

The Quantum Dictionary pattern~\cite{Gonciulea2019}, can be used to implement such an encoding. The original purpose of the Quantum Dictionary was to analyze the values of a given function (in particular, a polynomial) in an optimization context.

In this paper, we present a new usage, where the function is designed to map selected states to a single value. This function can be interpreted in multiple ways:

\begin{enumerate}  
    \item As a partition of all possible outcomes that exclusively places the selected ones into the same partition class.
    \item As a random variable on all possible outcomes of a quantum state.
    \item As a Hamiltonian that exclusively assigns the selected outcomes the same energy level. In particular, an Ising model can be efficiently encoded.
\end{enumerate} 
The last interpretation is especially useful in problems that can be naturally formulated in terms of Hamiltonians.

With the example of the Bell State, we have two sets of outcomes: those where the two digits match, and those where they do not. 
We can use a Quantum Dictionary with $2$ key qubits and $1$ value qubit, and the following function:
\begin{eqnarray}
f(x_0, x_1)= x_0 - x_1
\label{eq:state-encode-bell}
\end{eqnarray}
to represent the partition of the outcomes into the disjoint sets defined above.


\section{\label{sec:examples}Experimental Results on Combinatorial Problems}

In this section, we will apply the concepts described in the previous sections to practical problems that can be expressed in terms of Ising models. We ran the different property-encoding methods on real quantum hardware, the Honeywell System Model HØ quantum computer with quantum volume 64~\cite{QuantumVolume}. For all the experiments executed on the Honeywell quantum computer, we obtained the expected results with just one execution consisting of 1024 shots, without any calibration or noise mitigation.

\subsection{\label{subsec:zero-sum-subsets}Zero-Sum Subset Problem}

As mentioned in Section~\ref{subsec:state-encoding}, we can use the canonical oracle encoding to efficiently address the sum subset problems, represented by Equation~\ref{eq:sum-of-digits}. In particular, the \emph{zero-sum subset problem} is of interest---i.e. given a set of integers, we want to know if there is a non-empty subset whose elements add up to $0$. More precisely, the problem applies to sets with repeated elements, also called \emph{multisets}, which are modeled as arrays or lists in Computer Science.

For example, assume we want to find a subarray of $[2, 1, -5, 2]$ such that:
\begin{eqnarray}
S(x_0, x_1, x_2, x_3) = 2x_0 + x_1 - 5x_2 + 2x_3 = 0
\label{eq:zero_subset_sum_example}
\end{eqnarray}
where $x_i \in \{0,1\}$ for $i \in \{0, 1, 2, 3\}$. Using the Quantum Dictionary pattern, we can encode  equation~\ref{eq:zero_subset_sum_example} into the state of two quantum registers.

\begin{figure}[ht]
\includegraphics[width=7cm]{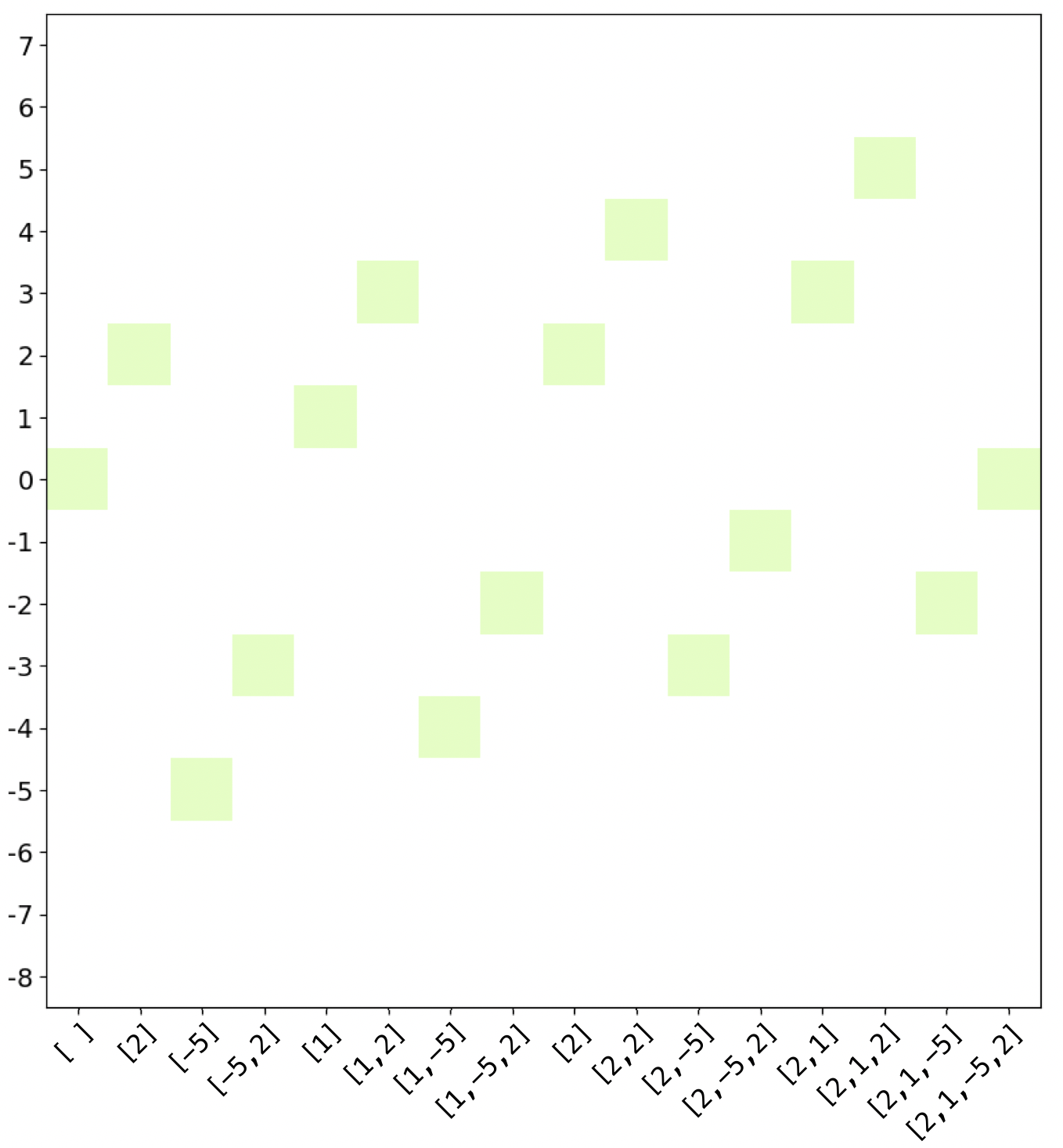}
\caption{\label{fig:zero_sum_pixel}A pixel graph visualization of a quantum state obtained by running the computation on a quantum simulator and representing the zero-sum subarray problem described in Equation~\ref{eq:zero_subset_sum_example}.} 
\end{figure}

This state is represented in Figure~\ref{fig:zero_sum_pixel}, in the form of a pixel-based visualization of the resulting quantum state vector. The input of the function is shown on the horizontal axis, with the corresponding values on the vertical axis. The hue of each pixel is determined by the phase of the amplitude, and the intensity of the color is determined by the magnitude. A full description of this visualization methodology can be found in Appendix~\ref{sec:pixel_based_visualization}.

From here, we can apply Quantum Counting with a simple oracle---one that matches $0$ in the value register---which will give us the number of subarrays whose elements add up to $0$. Note that the count includes the empty set, which by convention has zero sum.

\subsection{\label{subsec:calculating-fibonacci-numbers}Calculating Fibonacci Numbers}
In this section, we show how all three types of encoding can be applied on a simple Ising problem, chosen to model the computation of the well known Fibonacci numbers.

\subsubsection{\label{subsec:amplitude-encoding-examples}Heuristic Encoding}
Heuristic encoding allows us to classically count the total number of possible measured outcomes, i.e. those with a non-zero probability.

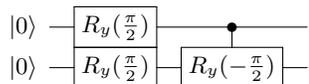
\begin{figure}[ht]
\begin{equation*}
    \Qcircuit @C=1.0em @R=0.0em @!R {
	 	\lstick{\ket{0}} & \gate{R_y(\frac{\pi}{2})} & \ctrl{1} & \qw\\
	 	\lstick{\ket{0}} & \gate{R_y(\frac{\pi}{2})} & \gate{R_y(-\frac{\pi}{2})} & \qw\\
	 }
\end{equation*}
\caption{\label{fig:consecutive-ones-circuit}A two-qubit circuit that eliminates the basis state $\ket{11}$ from the possible measurement outcomes.} 
\end{figure}

Let us consider a two-qubit quantum state where all outcomes are possible except $\ket{11}$. Heuristically, we obtain this state if the first qubit is in equal superposition, and the second is in equal superposition only when the first is measured as $0$. This can be done using $R_Y$ rotations, as shown in in Figure~\ref{fig:consecutive-ones-circuit}. An $R_X$ rotation will also work, and depending on the hardware, may be more efficient.

\begin{figure}[ht]
\includegraphics[width=8cm]{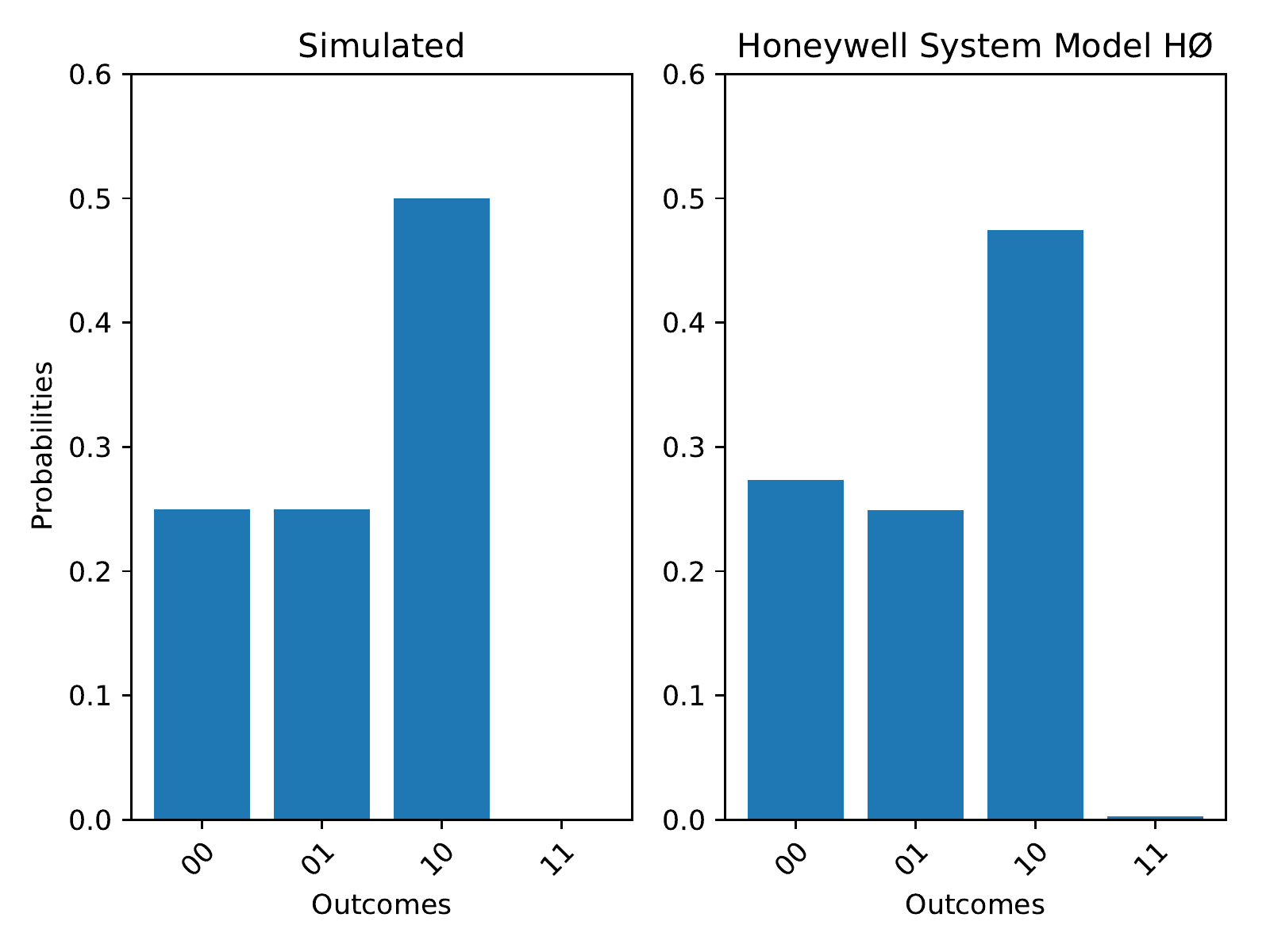}
\caption{\label{fig:consecutive-ones-example}The result of running the circuit from Figure~\ref{fig:consecutive-ones-circuit} on a simulator (left) and the Honeywell System Model HØ quantum computer with 1024 shots (right).} 
\end{figure}

\begin{figure}[ht]
\includegraphics[width=8cm]{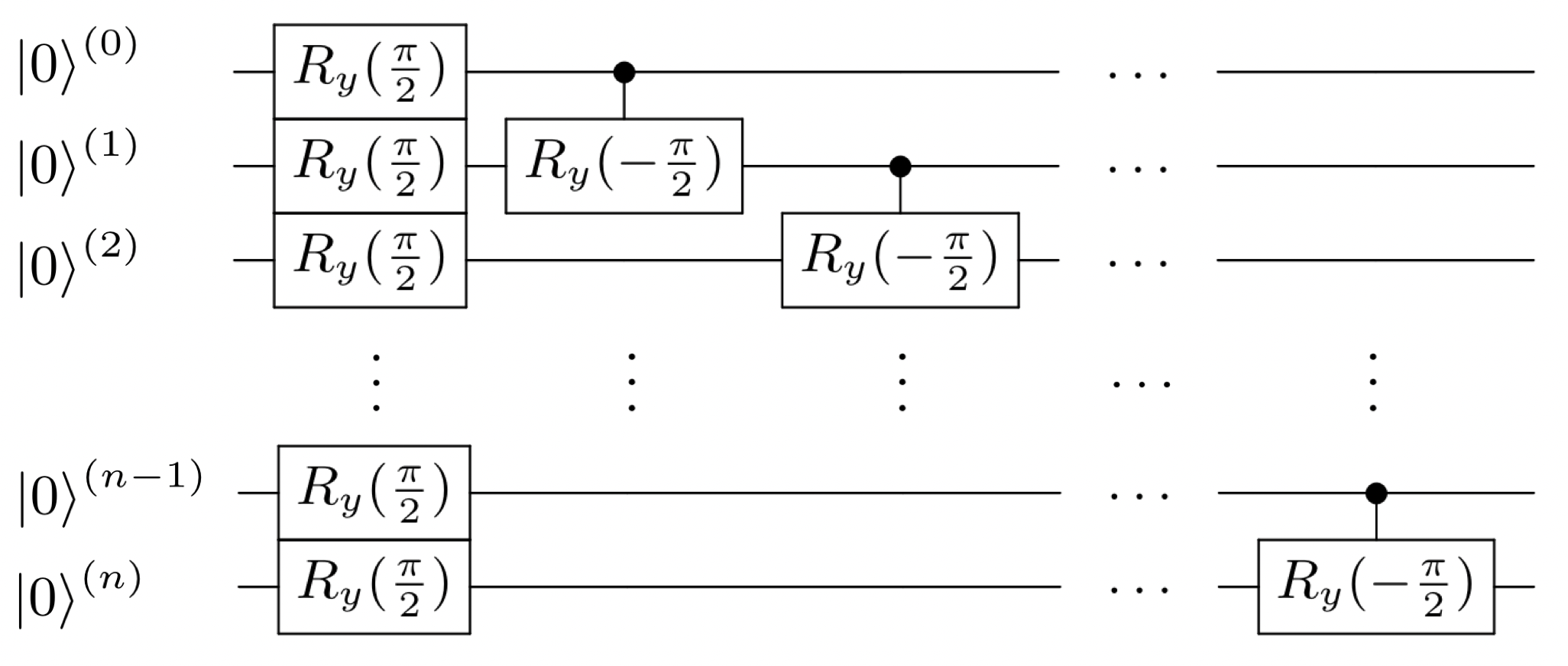}
\caption{\label{fig:generalized-consecutive-one-circuit}The generalized form of the circuit shown in Figure~\ref{fig:consecutive-ones-circuit}, which generates all possible outcomes without consecutive ones.} 
\end{figure}

The corresponding probability distribution is shown in Figure~\ref{fig:consecutive-ones-example}. We can generalize this circuit to $n$ qubits to generate outcomes without consecutive $1$s, as shown in Figure~\ref{fig:generalized-consecutive-one-circuit}.

We denote by $F(n)$ the number of binary strings of length $n$ without consecutive $1$s. With this notation, we get $F(1)=2$, $F(2)=3$, $F(3)=5$, $F(4)=8$, $F(5)=13$, etc. Note that each term in this sequence is the sum of the previous two, a property of the Fibonacci numbers that is often used in classical computations to make their calculation efficient. Also note that in some literature the first two Fibonacci numbers in the sequence are both $1$, but in our notation indices represent the length of the binary strings, so the sequence starts from $2$.

\begin{figure}[ht]
\includegraphics[width=8cm]{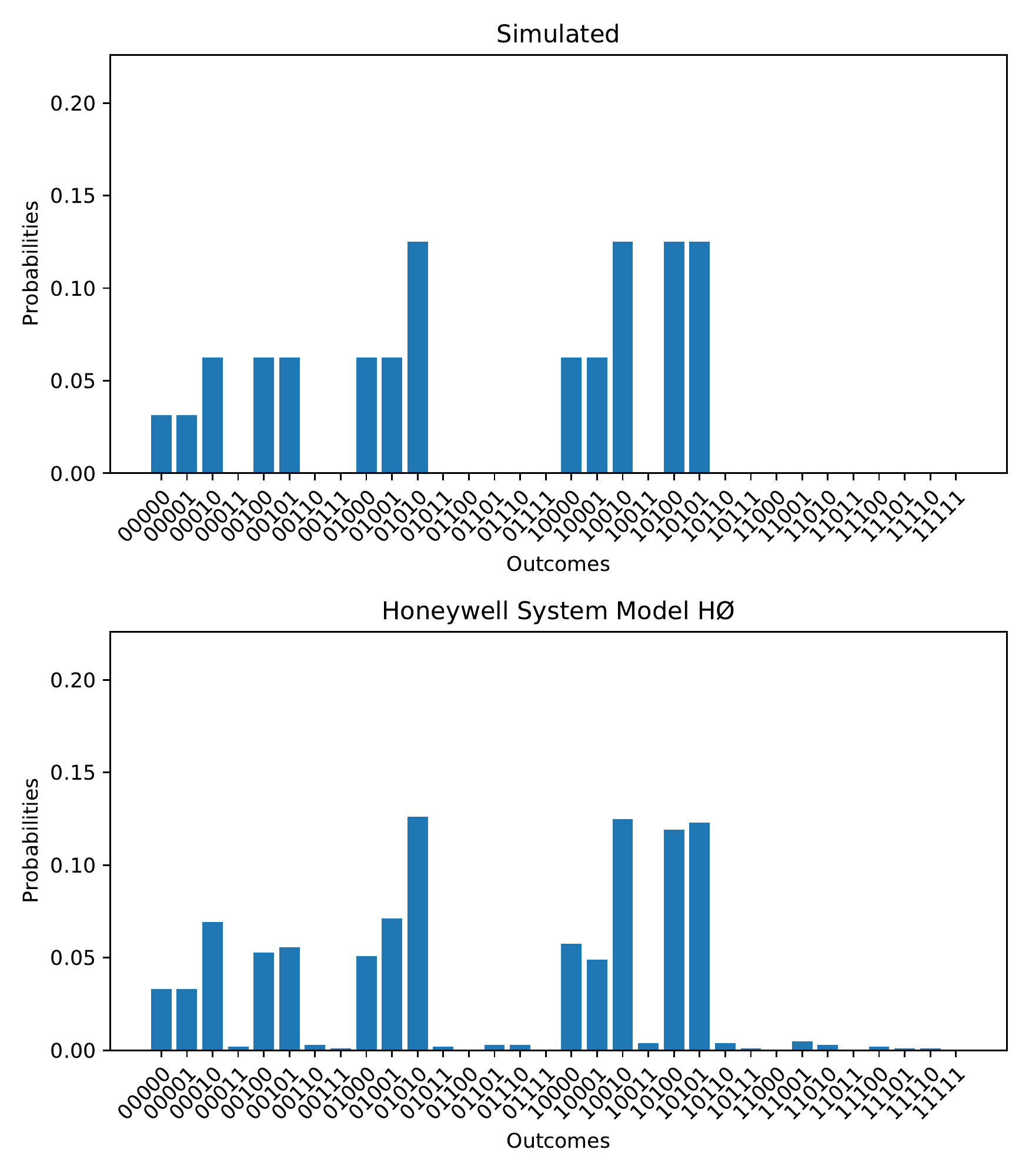}
\caption{\label{fig:brute_force_fib_results}The result of running the circuit from Figure~\ref{fig:generalized-consecutive-one-circuit} for $n=5$ qubits on a simulator (top) and the Honeywell system model HØ quantum computer with 1024 shots (bottom).} 
\end{figure}

For example, we can calculate $F(5)$ using the circuit in Figure~\ref{fig:generalized-consecutive-one-circuit} for $n=5$. The results are seen in Figure~\ref{fig:brute_force_fib_results}.

\subsubsection{\label{subsec:oracle-encoding-examples}Naive Oracle Encoding}

\begin{figure}[ht]
\includegraphics[width=7cm]{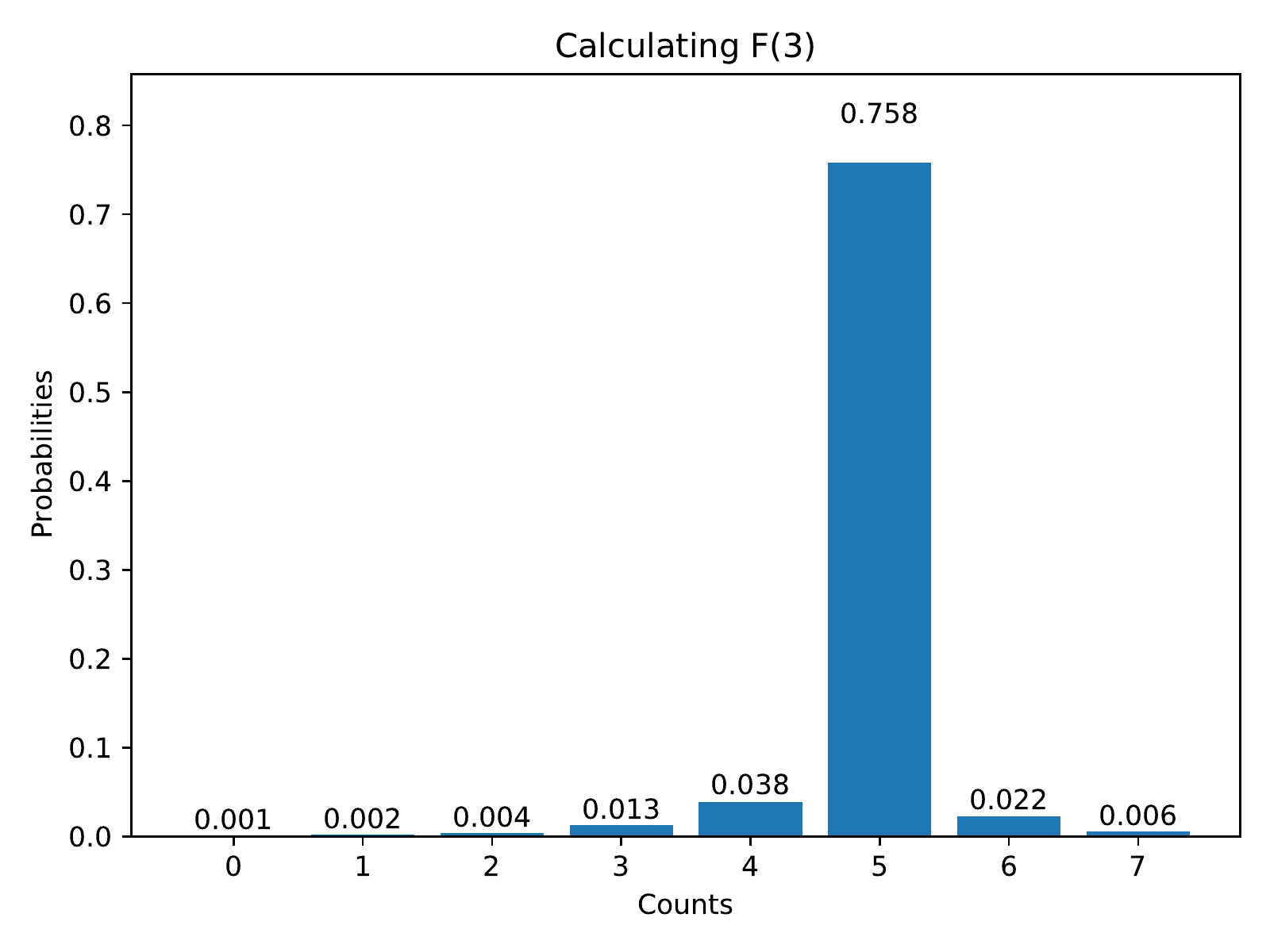}
\caption{\label{fig:fib_count_3_qubits}The simulated result of a Quantum Counting circuit using an oracle that recognizes outcomes without consecutive $1$s.} 
\end{figure}

We can also calculate Fibonacci numbers using the naive oracle encoding approach, by providing an oracle that selects binary strings with no consecutive $1$s. For example, in Python we can specify the oracle's predicate as the following function:

\begin{verbatim}
    predicate = lambda k: k & k >> 1 == 0
\end{verbatim}

In this example, all possible outcomes are placed in equal superposition. A Quantum Counting circuit is then applied, using the oracle described above. The result can be seen in Figure~\ref{fig:fib_count_3_qubits}.

In contrast to the heuristic encoding, which requires sampling across multiple runs, Quantum Counting is designed to minimize the number of times we repeat the computation to obtain the desired result.

\subsubsection{\label{subsec:state-encoding-examples}Canonical Oracle Encoding}

Given an $n$-qubit quantum system, we can mathematically represent the number of pairs of consecutive $1$s in a possible outcome (a binary strings of length $n$)  as:
\begin{eqnarray}
H(x_0, x_1, ..., x_{n-1}) = \sum_{i=0}^{n-2}  x_ix_{i+1}
\label{eq:no-consecutive-ones-equation}
\end{eqnarray}
where $x_i \in \{0,1\}$ for $i \in \{0, 1, \ldots, n-1\}$. The property of an outcome $(x_0, x_1, ..., x_{n-1})$ having no consecutive $1$s translates into $H(x_0, x_1, ..., x_{n-1})=0$.

As mentioned in Section~\ref{subsec:amplitude-encoding-examples}, the $n$th Fibonacci number $F(n)$ is defined as the number of the binary strings of length $n$ with no consecutive $1$s. 
The function in the right-hand side of Equation~\ref{eq:no-consecutive-ones-equation} also has the form of an Ising model, and $0$ is its minimum value. This is in fact an encoding of the Hamiltonian that associates to each possible configuration the number of consecutive $1$s in its binary representation.

As an example, a representation of this encoding for $n=3$ can be seen in Figure~\ref{fig:state_encoding_fib}, using the visualization technique described in Appendix~\ref{sec:pixel_based_visualization}. Just looking at the visualization, we see that there are 5 states with zero sum. Therefore, $F(3)=5$. Note that the Quantum Counting procedure would follow this encoding, and was not included in the run on the real computer.
When the Quantum Counting procedure is run on a simulator, the result is similar to Figure~\ref{fig:fib_count_3_qubits}.


\begin{figure}[ht]
\includegraphics[width=7cm]{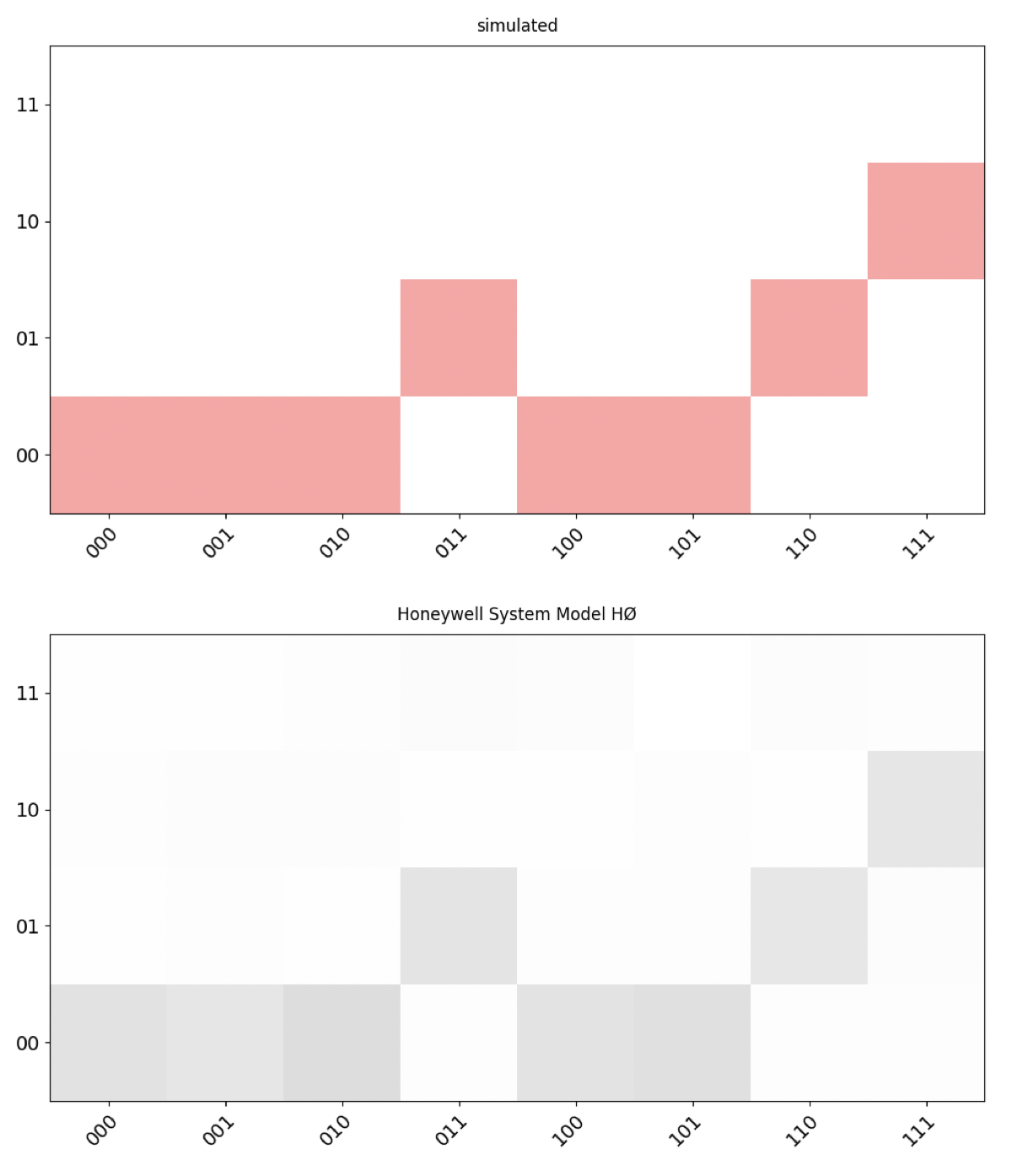}
\caption{\label{fig:state_encoding_fib}Two pixel graph visualizations of a quantum state corresponding to the function in Equation~\ref{eq:no-consecutive-ones-equation} for $n=3$, using a simulator (top) and the Honeywell System Model HØ quantum computer (bottom) with 1024 shots.} 
\end{figure} 

This method uses the Generalized Amplitude Amplification/Estimation procedure described in Section~\ref{sec:smart-oracles}, where the operator $B$ is the function encoding, and the oracle is simply matching $\ket{0}$ in the value register.

\section{\label{sec:conclusion} Conclusion}
In this paper, we have shown a canonical way to construct a quantum oracle, starting from an algebraic expression that describes a set of selected basis states.
We introduced a general form of the Grover iterate where the oracle is of the form $O=B^{\dagger}O_BB$, with $B$ being a property-encoding operator that maps the selected states to a single function value, and $O_B$ is a binary-matching oracle.
This general form creates a foundation for further optimizations of Quantum Search.

We investigated practical approaches to apply canonical oracles to concrete examples of NP-hard problems, such as the zero-sum subset problem, and compared heuristic, naive oracle, and canonical oracle encodings in the context of a simple Ising Hamiltonian that models the computation of Fibonacci numbers.
We have also made a connection between concepts from classical probabilistic spaces, namely events and random variables, with the corresponding quantum computing constructs (Born Rule, Amplitude Estimation, and Quantum Dictionary).
We validated our results on the Honeywell System Model HØ quantum computer (quantum volume 64).

\section*{Acknowledgements}
We would like to thank the \textbf{Honeywell Quantum Solutions} team, in particular \textbf{Tony Uttley}, \textbf{Brian Neyenhuis}, and \textbf{David Hayes}, for their invaluable help on the execution of our experiments on the Honeywell quantum computer.

\textbf{Honeywell} was gracious enough to share the results of their quantum volume 64 evaluation, prior to their upcoming publishing of those results.

Special thanks to \textbf{Denise Ruffner} and the \textbf{Cambridge Quantum Computing} team for giving us access to their t|ket> library for some of our initial runs on the Honeywell quantum computer.

\section*{Disclaimer}
This paper was prepared for information purposes by the Future Lab for Applied Research and Engineering (FLARE) Group of JPMorgan Chase \& Co. and its affiliates, and is not a product of the Research Department of JPMorgan Chase \& Co. 
JPMorgan Chase \& Co. makes no explicit or implied representation and warranty, and accepts no liability, for the completeness, accuracy or reliability of information, or the legal, compliance, tax or accounting effects of matters contained herein.
This document is not intended as investment research or investment advice, or a recommendation, offer or solicitation for the purchase or sale of any security, financial instrument, financial product or service, or to be used in any way for evaluating the merits of participating in any transaction.

\section*{Appendix}
\appendix

\section{\label{sec:concepts}Fundamental Concepts}

In this section, we recall some fundamental concepts of classical probability theory and quantum computing in the context of properties that define events and good quantum states.

\subsection{Classical Probability Spaces}
A \emph{classical probability space} $(S, p)$ consists of a finite set $S$ of \emph{outcomes}, and a \emph{probability function} $p: S \rightarrow [0,1] \label{eq:probability-function}$ obeying the following \emph{normalization property}:
 
\begin{eqnarray}
\sum_{s \in S} p(s) = 1
\label{eq:sum-of-probability}
\end{eqnarray}

An arbitrary subset $E \subseteq S$ is called an \emph{event}, and its probability is defined as:

\begin{eqnarray}
p(E) = \sum_{s \in E} p(s)
\end{eqnarray}

A typical way to define the outcomes of an event $E$ is by specifying a property that uniquely characterizes the outcomes in $E$. 
For example, a $6$-sided die landing on an even number is an event $E$ in the set of outcomes $S = \{1,2,3,4,5,6\}$. The property that uniquely characterizes $E$ is that each of its elements must be an even number.  Therefore, $E = \{2,4,6\}$. 

\subsection{Quantum State}
In quantum computing, the \emph{state} of an $n$-qubit quantum system can be interpreted as a mapping $a: S \rightarrow \mathbb{C}$, where $S = \{0, 1, \ldots, 2^n-1\}$, satisfying the following \emph{normalization property}:

\begin{eqnarray}
\sum_{s \in S} \lvert a(s) \rvert^2 = 1
\label{eq:quantum-normalization-property}
\end{eqnarray}

Typically, a quantum state is expressed using the ket notation~\cite{Dirac1939ANN}, as follows:

\begin{eqnarray}
\sum_{s = 0}^{2^n-1} a(s) \ket{s}
\label{eq:ket}
\end{eqnarray}

\subsection{The Born Rule}
The Born Rule~\cite{Mermin2007} makes a quantum state tangible, by defining what happens when the state of a quantum system is measured.
Specifically, it establishes that any measurement results in a single outcome, and the probability of measuring a particular outcome is the squared magnitude of the amplitude corresponding to that outcome.

This was a intelligent guess on Born's part, without a explicit justification. There have been multiple efforts to derive the Born Rule from more fundamental principles~\cite{Ball2019}.

The interpretation of what the probabilities represent is the part of the Born Rule that is debated by physicists and philosophers. What is mathematically indisputable is that the state of an $n$-qubit quantum system specified by $a: S \rightarrow \mathbb{C}$, where $S = \{0, 1, ..., 2^n-1\}$, gives rise to a classical probabilistic space whose probability function $p: S \rightarrow [0, 1]$ is defined as follows: 

\begin{eqnarray}
p(s) = \lvert a(s) \rvert^2
\label{eq:born-rule}
\end{eqnarray}

Note that sometimes it is useful to interpret $S$ as the set of binary strings of length $n$.

The probability of a set of basis states $E \subseteq S$ is obtained as follows:

\begin{eqnarray}
p(E) = \sum_{s \in E} \lvert a(s) \rvert^2
\label{eq:born-rule-events}
\end{eqnarray}

As in the classical probabilistic context, a set of basis states (or outcomes) is typically specified by a property that uniquely characterizes them.  In popular literature  \cite{Brassard2000}, the states that satisfy a given property are referred to as \emph{good} or {eligible} with respect to that property, and those that do not as \emph{bad} or \emph{ineligible}. The most common way to use such properties is through oracles.

\subsection{Oracles\label{subsec:oracles}}

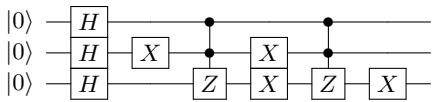
\begin{figure}[ht]
\begin{equation*}
    \Qcircuit @C=1.0em @R=0.0em @!R {
	 	\lstick{\ket{0}} & \gate{H} & \qw      & \ctrl{1}    & \qw      & \ctrl{1}    & \qw      & \qw\\
	 	\lstick{\ket{0}} & \gate{H} & \gate{X} & \ctrl{1}    & \gate{X} & \ctrl{1}    & \qw      & \qw\\
	 	\lstick{\ket{0}} & \gate{H} & \qw      & \gate{Z}    & \gate{X} & \gate{Z}    & \gate{X} & \qw\\
	 }
\end{equation*}
\caption{\label{fig:dynamic-oracle-example}The circuit diagram for a dynamic oracle that recognizes the elements in $\{5, 6\}$.}
\end{figure}

In the context of quantum computing, an \emph{oracle} is an important subroutine that marks the basis states that satisfy a given property. Typically, an \emph{oracle} uses a classically-defined predicate derived from a property. The desired states are typically matched using a series of CNOT gates, which can become expensive--particularly if the oracle is applied several times throughout the course of an algorithm, as in the case of Grover's Search. As an example, consider the oracle circuit shown in Figure~\ref{fig:dynamic-oracle-example}, which selects all the states that correspond to the elements of $\{5, 6\}$.

\section{\label{sec:amplitude-estimation}Amplitude Estimation}

Suppose we have an $n$-qubit quantum system, and let $S = \{0, 1, ..., 2^n-1\}$.
Assume we are given a unitary operator $A$ that prepares the quantum state:

\begin{eqnarray}
A\ket{0}_n = \sum_{s = 0}^{2^n-1} a(s) \ket{s}
\label{eq:a-operator}
\end{eqnarray}

Furthermore, assume we are given a predicate function $f: S \rightarrow \{0, 1\}$, and let $E$ be the set of basis states that are mapped to $1$; $E$ is the set of the good states.
Then, the Amplitude Estimation algorithm provides a way to efficiently calculate the probability:

\begin{eqnarray}
p(E) = \sum_{s \in E} \lvert a(s) \rvert^2
\label{eq:ae-prob-of-event}
\end{eqnarray}

In the following subsections, we describe each component of Amplitude Estimation, along with the relevant implementation details.

\subsection{The Grover Iterate\label{subsec:grover-iterate}}

The \emph{Grover iterate} $G$ is comprised of three components~\cite{Brassard2000}:
\begin{enumerate}  
\item A unitary \emph{state-preparation operator} $A$
\item An \emph{oracle} $O$ corresponding to a predicate $f$, which recognizes all the elements in $E$ and multiplies their amplitudes by $-1$. For example, if $n$ is the number number of qubits:
\begin{eqnarray}
OA\ket{0}_n = \sum_{s \notin E} a(s)\ket{s}_n - \sum_{s \in E} a(s)\ket{s}_n
\end{eqnarray}
\item The \emph{diffusion operator} $D$, which multiplies the amplitude of the $\ket{0}_n$ state by $-1$
\end{enumerate}
Given the above, Grover iterate is defined as  $G = -ADA^{\dagger}O$. For the normalized states defined as follows:
\begin{eqnarray}
\ket{X_0} = \frac{1}{\sqrt{p(E{'})}}\sum_{s \in E{'}} a(s)\ket{s}
\label{eq:x0}
\end{eqnarray}
\begin{eqnarray}
\ket{X_1} = \frac{1}{\sqrt{p(E)}}\sum_{s \in E} a(s)\ket{s}
\label{eq:x1}
\end{eqnarray}
where $E{'}=S \setminus E$ is the complement of $E$, after applying $k$ iterations of $G$ to $A\ket{0}$, we obtain:
\begin{equation}
\begin{aligned}[b]
G^{k}A\ket{0} &= \cos{((2k+1)\theta)}\ket{X_0} + \\
&\quad\sin{((2k+1)\theta)}\ket{X_1}
\label{eq:k-applications-grover}
\end{aligned}
\end{equation}
where $\theta$ is the angle in $[0, \pi]$ that satisfies $\sin^2{\frac{\theta}{2}}=p(E)$.

\subsection{Grover's Search Algorithm}

For an integer $k \geq 0$, measuring the state described in Eq.~\ref{eq:k-applications-grover} will return an eligible state $s \in E$ with probability equal to $\sin^2{((2k+1)\theta)}$. 
Choosing $k = \lfloor\frac{\pi}{4}\sqrt\frac{|S|}{|E|}\rfloor$ will maximally amplify the amplitudes of the eligible states~\cite{Grover1996}. 
If $|E|$ is not known, $G$ can be applied a random number of times in an adaptive manner~\cite{Durr1996, Bulger2003, Baritompa2005, Gilliam2019}. 

\subsection{Amplitude Estimation Algorithm}

\begin{figure}[ht]
\includegraphics[width=8cm]{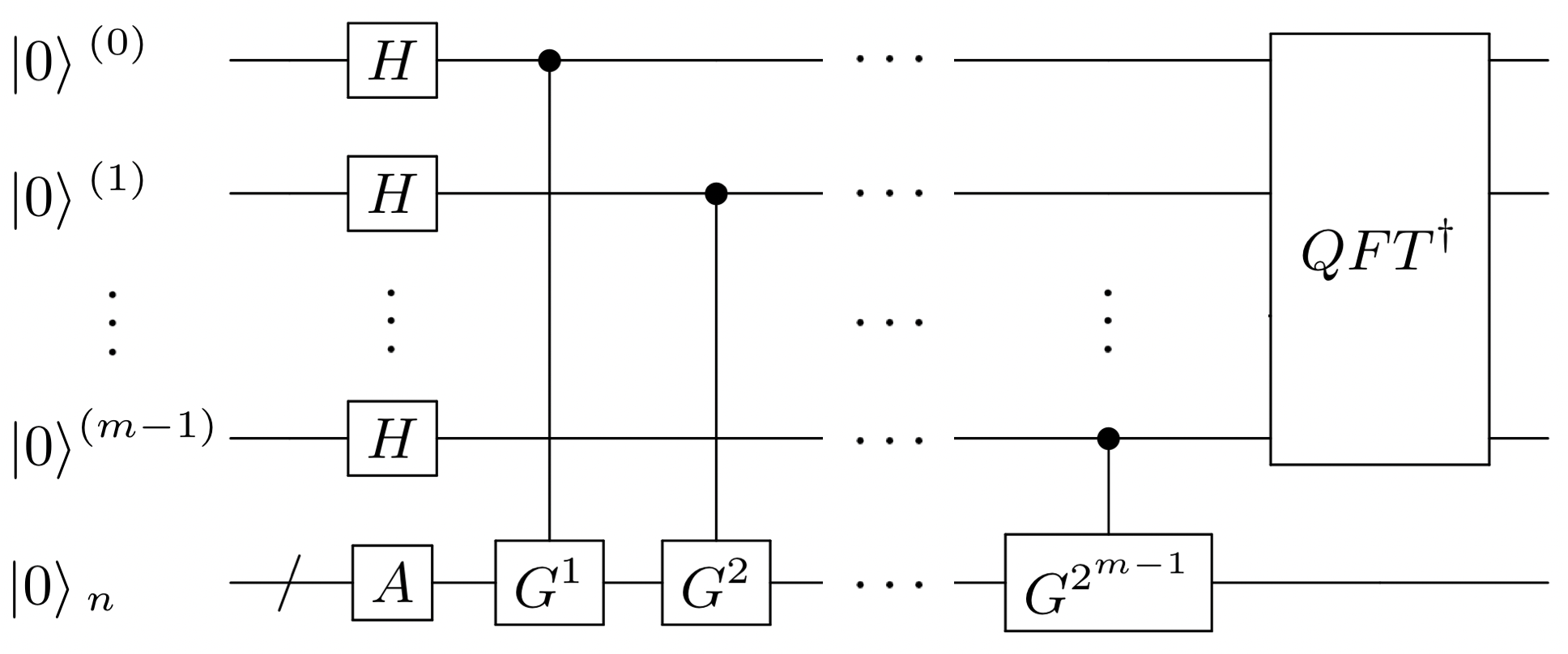}
\caption{\label{fig:ae-circuit}A circuit diagram for the Amplitude Estimation Algorithm. The main register consists of $n$ qubits, and the results register consists of $m$ qubits.} 
\end{figure}

Given a state preparation operator $A$ and an oracle $O$ that recognizes a set of basis states $E$, we construct the Grover iterate $G$ as described in Appendix~\ref{subsec:grover-iterate}. The Amplitude Estimation algorithm consists of the following steps, shown as a circuit in Figure~\ref{fig:ae-circuit}.

 \begin{enumerate}
    \item We initialize a circuit consisting of two registers. The first register consists of $n$ qubits. The second register will encode the estimate and consists of $m$ qubits, where $m$ depends on the desired precision.
    \item We apply the state preparation operator $A$ to the first register, and a Hadamard gate to each qubit in the second register.
    \item For each result qubit $k \in \{0, 1, ..., m-1\}$, we apply $G^{2^k}$ to the first register, controlled on result qubit $k$. \label{ae-steps:apply-grover}
    \item Apply the inverse Quantum Fourier Transform to the result register.
    \item We measure the state of the result register as an integer $y \in \{0, 1, ..., 2^m - 1\}$, which translates into an estimate $\sin^{2}{(y\pi/2^m)}$ of $p(E)$. We get one of the best two estimates with a probability of $\frac{8}{\pi^2}$ or greater~\cite{Brassard2000}. \label{ae-steps:classical-processing}
\end{enumerate}

In practice, we can ignore the negative sign in $G$ and use $\cos^2$ instead of $\sin^2$ in Step~\ref{ae-steps:classical-processing}. As introduced in \cite{Suzuki2019}, we can also remove the need for $QFT^{\dag}$. In certain literature, the state preparation operator is expected to identify outcomes that satisfy a given property with an ancillary qubit (which is then matched by a simple oracle), but this is unnecessary.


\subsection{Quantum Counting}
Typically Quantum Counting is presented as a consequence of Amplitude Amplification, where the operator $A$ prepares an equal superposition state in the first register, and the oracle $O$ recognizes the set of states that need to be counted. The count is just the probability estimated in the previous section, multiplied by the total number of possible states of the first register $2^n$.

However, more sophisticated setups are possible, like the one in the Quantum Dictionary pattern~\cite{Gonciulea2019}, where the first register is replaced by two (let us call them key and value) entangled registers, with oracles that select states based on the value register. 

We will provide examples of both simple and more complex applications of Quantum Counting.

\section{\label{sec:quantum_dictionary} Quantum Dictionary}
 
The \emph{Quantum Dictionary} was introduced in~\cite{Gonciulea2019} as a quantum computing pattern for encoding functions, in particular polynomials, into a quantum state using geometric sequences. The paper shows how quantum algorithms like search and counting applied to a quantum dictionary allow to solve combinatorial  optimization and QUBO problems more efficiently than using classical methods. We summarize the construction in the following text. 
 
Given an $m$-qubit register and an angle $\theta \in [-\pi, \pi)$, we wish to prepare a quantum state whose state vector represents a "periodic signal" equivalent to a geometric sequence of length $2^m$: 
\begin{eqnarray}
G(\theta) = (1, e^{i\theta}, \ldots, e^{i(2^m - 1)\theta}).
\end{eqnarray}

In other words, after normalizing, we need a unitary operator defined by:
\begin{eqnarray}
U_{\text{G}}(\theta) H^{\otimes m}\ket{0}_m = \frac{1}{\sqrt{2^m}}\sum_{k=0}^{2^m - 1} e^{ik\theta} \ket{k}_m.
\end{eqnarray}

\begin{figure}[ht]
    \mbox{
    \Qcircuit @C=1em @R=0em @!R {
    0 & {} & {} & {} & \gate{R(2^{m - 1}\theta)} & \qw & \qw  \\ 
    \vdots & & & & \ldots \\
    m-1-i & {} & {} & {} & \gate{R(2^{i}\theta)} & \qw & \qw  \\ 
    \vdots  & & & & \ldots \\
    m - 1 & {} & {} & {} & \gate{R(\theta)} & \qw & \qw   \\ 
    }
}
\caption{\label{fig:geom}The $U_{\text{G}}(\theta)$ circuit.}
\end{figure}
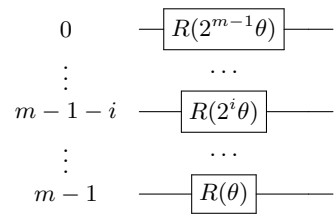

The simplest implementation of $U_{\text{G}}(\theta)$ uses the phase gate $R(\theta)$ that, which when applied to a qubit, rotates the phase of the amplitudes of the states having $\ket{1}$ in that qubit. In Qiskit, this gate is the $U_1(\theta)$ operator \cite{qiskit}. 
The circuit for $U_{\text{G}}(\theta)$ is shown in Fig.~\ref{fig:geom}, and consists of applying the gate $R(2^{i}\theta)$ to the qubit $m - 1 - i$ in the $m$-qubit register prepared in a state of equal superposition.

Given an integer $-2^{m - 1} \le k < 2^{m - 1}$, if we apply $U_{\text{G}}(\frac{2\pi}{2^{m}}k)$, followed by the inverse QFT to an $m$-qubit register prepared in a state of equal superposition, we end up with ${k \pmod{2^m}}$ being encoded in the register, as shown in Fig.~\ref{fig:geom-qft-integer}. 

\begin{figure}[ht]
    \mbox{
    \Qcircuit @C=1em @R=1em {
    & \lstick{H^{\otimes m}\ket{0}_m{}} &
    \gate{U_{\text{G}}(\frac{2\pi}{2^{m}}k)}  & \gate{QFT^\dag} & \qw & 
    = & & & & &  \ket{k \pmod{2^m}}_m{} \\ 
    }
}
\caption{\label{fig:geom-qft-integer}Geometric sequence encoding of an integer $k$.}
\end{figure}

This representation is called the binary Two's Complement of $k$, which just adds $2^m$ to negative values $k$, similar to the way we can represent negative angles with their complement, e.g. equating $-\pi/4$ with $7\pi/4$.
The reason this representation occurs naturally in this context is due to the fact that rotation composition behaves like modular arithmetic.

It is worth mentioning that there is an alternative method described in~\cite{Gonciulea2019}, which uses the $R_y$ family of gates with the eigenstate $1/\sqrt{2}(i\ket{0} + \ket{1})$, independent of the rotation angle.

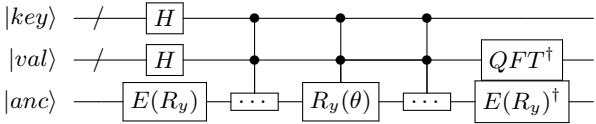
\begin{figure}[ht]
    \mbox{
    \Qcircuit @C=1em @R=0em @!R {
    & \lstick{\ket{key}{}} & \qw{/} & \gate{H} &\ctrl{2} & \ctrl{2} & \ctrl{2} & \qw & \qw  \\
    & \lstick{\ket{val}{}} & \qw{/} & \gate{H} &\ctrl{1} & \ctrl{1} & \ctrl{1} \qw & \gate{QFT^\dag}  & \qw   \\ 
    & \lstick{\ket{anc}{}} & \qw & \gate{E(R_y)} & \gate{\text{\ldots}} & \gate{R_y(\theta)} & \gate{\text{\ldots}} & \gate{E(R_y)^\dag} & \qw   \\ 
    }
}
\caption{\label{fig:encoding-initial-state}Quantum Dictionary circuit.}
\end{figure} 
 
The $R_y$ gate is applied to an ancillary register containing one of its eigenstates prepared by an operator $E(R_y)$, conditioned on both the key and value registers. The rotation angle $\theta$ is different for each application, representing a number that contributes to the values corresponding to keys that have the conditioned key as a subset. In particular, when encoding a polynomial, a rotation will be applied for each of its coefficients.

\begin{figure}[ht]
    \mbox{
    \Qcircuit @C=1em @R=1em {
    & \lstick{\ket{0}{}} & \gate{R_x(\pi/2)} & \gate{Z} & \gate{X} & \qw & \push{\rule{.3em}{0em}=\rule{.3em}{0em}} & \gate{E(R_y)} & \qw \\ 
    }
}
\caption{\label{fig:eigenstate-preparation}Eigenstate preparation for $R_y$.}
\end{figure} 

\section{\label{sec:pixel_based_visualization} Pixel-Based Quantum State Visualization}

\begin{figure}[ht]
\includegraphics[width=3cm]{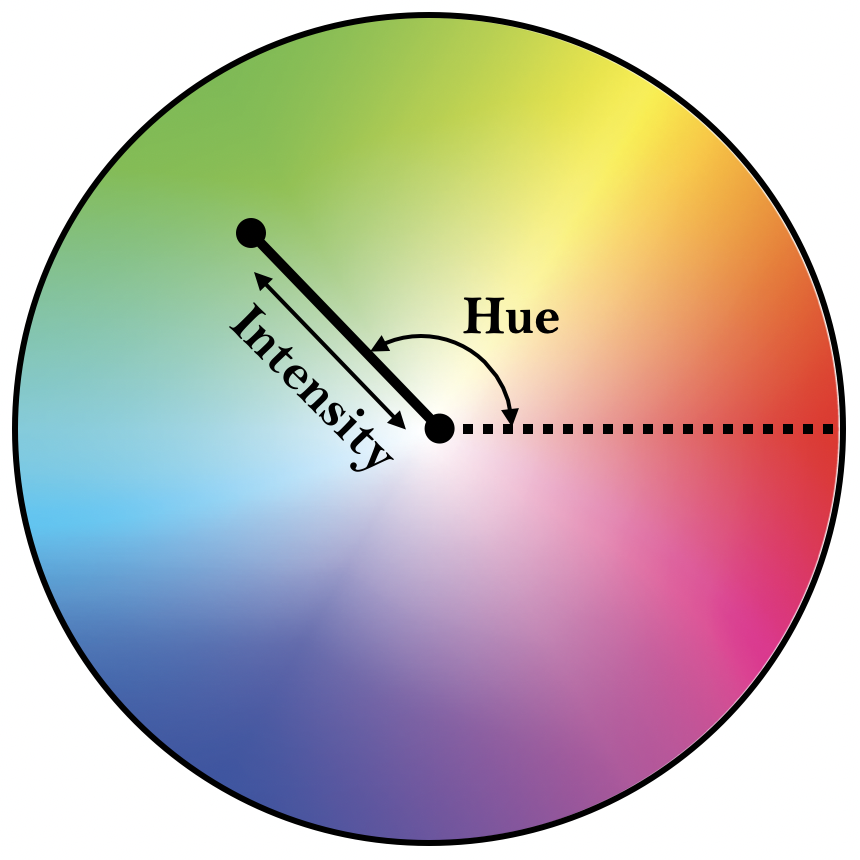}
\caption{\label{fig:color_wheel}A complex number represented in polar form, overlaid onto a color wheel. The phase of the amplitude determines the hue, and the magnitude determines the intensity.} 
\end{figure}

Amplitudes are complex numbers that have a direct correspondence to colors - mapping angles to hues and magnitudes to intensity - as seen in Figure~\ref{fig:color_wheel}.

\begin{figure}[ht!]
\includegraphics[width=8cm]{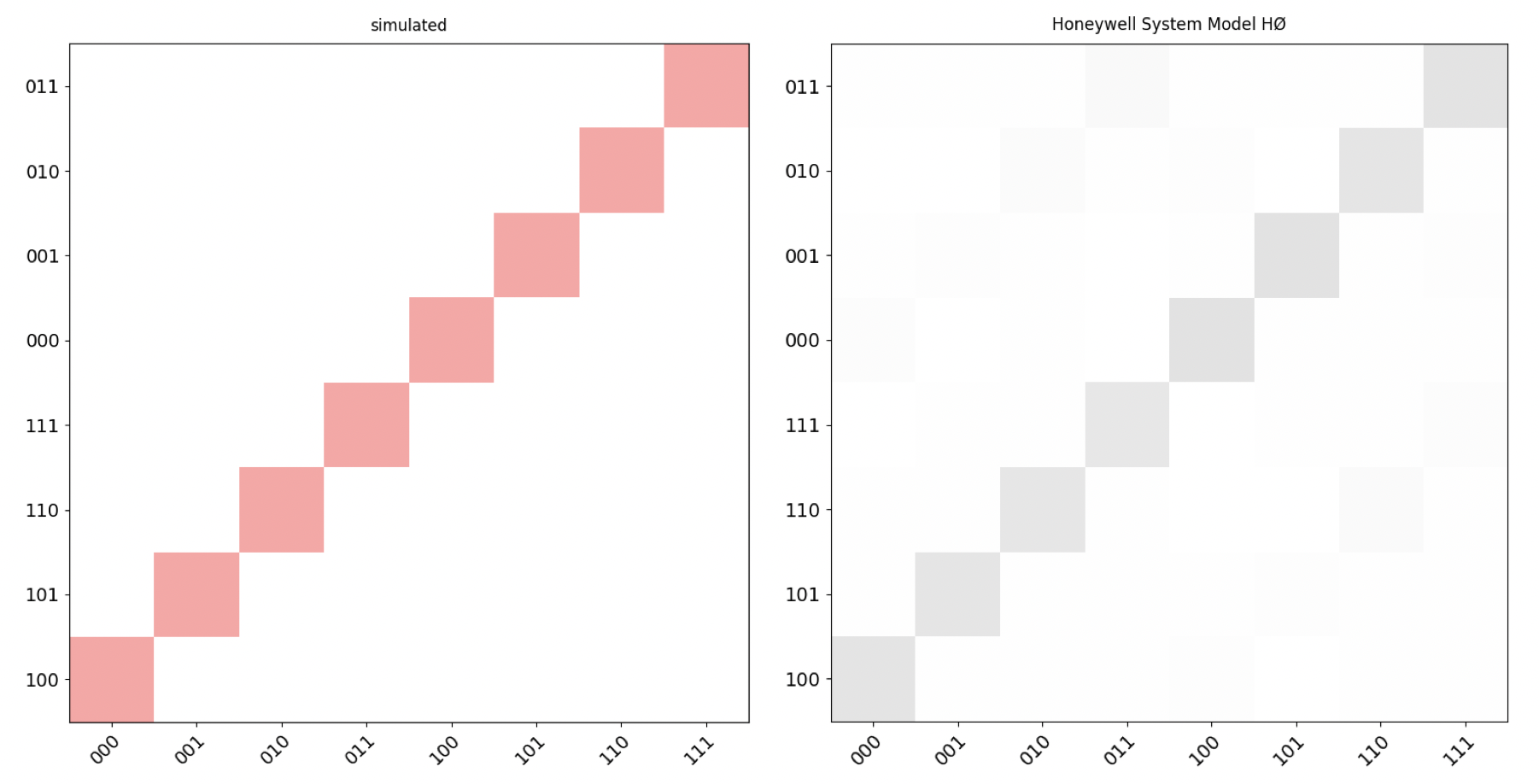}
\caption{\label{fig:pixel_graph} A pixel graph visualization of a quantum state (left), generated from a quantum computation that encodes the function $f(x)=x$ with 6 qubits (3 for the input, 3 for the output). We also show the results from the Honeywell System Model HØ quantum computer with 1024 shots (right). Note that the pixels are grey in the real case, as we don't know the phases of the amplitudes.} 
\end{figure} 

Using this technique, we can represent the quantum state as a column of pixels, where each pixel corresponds to its respective amplitude. If the computation contains two entangled registers, such as with a Quantum Dictionary, the visualization is also useful in a tabular form. For example, let us visualize a simple linear function - $f(x)=x$. As seen in Figure~\ref{fig:pixel_graph}, the inputs are represented by the key register (each column in the pixel graph), and the outputs by the value register (each row).

\FloatBarrier
\bibliography{main}
 
\end{document}